\RequirePackage{fix-cm}
\documentclass[twocolumn,natbib]{svjour3}          
\smartqed  
\setcitestyle{numbers}
\usepackage{lineno}
\modulolinenumbers[5]

\usepackage{scalefnt}
\usepackage{amssymb}
\usepackage{url}
\usepackage{amsfonts,amsmath}
\usepackage{graphicx}
\usepackage{stfloats}
\usepackage{subfigure}

\usepackage{multirow}
\usepackage{xfrac}

\usepackage{stmaryrd}
\usepackage{stfloats}
\usepackage{color}
\usepackage{caption}
\usepackage{listings}
\usepackage[table]{xcolor}

\usepackage{algorithm}
\usepackage{algpseudocode}
\usepackage{subfigure}
\usepackage{setspace} 

\usepackage[T1]{fontenc}

\usepackage{epstopdf} 

 \journalname{arXiv.org}

\begin{document}

\title{Real-time Neural Networks Implementation Proposal for Microcontrollers}


\author{Caio J. B. V. Guimar\~aes and Marcelo A. C. Fernandes}

\authorrunning{C.J.B.V. Guimar\~aes and M.A.C. Fernandes} 

\institute{Caio J. B. V. Guimar\~aes,  \at
              Laboratory of Machine Learning and Intelligent Instrumentation, Federal University of Rio Grande do Norte, Natal 59078-970, Brazil. \\
              ORCiD: 0000-0002-8392-5556\\
              \email{caio.b.vilar@gmail.com}  \\
              Marcelo A. C. Fernandes,  \at
              Department of Computer Engineering and Automation, Federal University of Rio Grande do Norte, 59078-970, Natal, RN, Brazil. \\
              ORCiD: 0000-0001-7536-2506\\
              \email{mfernandes@dca.ufrn.br}  \\
}

\date{Received: date / Accepted: date}

\maketitle

\begin{abstract}
The adoption of intelligent systems with Artificial Neural Networks (ANNs) embedded in hardware for real-time applications currently faces a growing demand in fields like the Internet of Things (IoT) and Machine to Machine (M2M). However, the application of ANNs in this type of system poses a significant challenge due to the high computational power required to process its basic operations. This paper aims to show an implementation strategy of a Multilayer Perceptron (MLP) type neural network, in a microcontroller (a low-cost, low-power platform).  A modular matrix-based MLP with the full classification process was implemented, and also the backpropagation training in the microcontroller. The testing and validation were performed through Hardware in the Loop (HIL) of the Mean Squared Error (MSE) of the training process, classification result, and the processing time of each implementation module. The results revealed a linear relationship between the values of the hyperparameters and the processing time required for classification, also the processing time concurs with the required time for many applications on the fields mentioned above. These findings show that this implementation strategy and this platform can be applied successfully on real-time applications that require the capabilities of ANNs.
\end{abstract}

\section{Introduction}

The microcontrollers ($\mu$Cs) have been applied in many areas: industrial automation, control, instrumentation, consumer electronics, and other various areas. Nonetheless, there is an ever-growing demand for these devices, especially in emerging sectors like the Internet of Things (IoT), Smart Grid, Machine to Machine (M2M) and Edge Computing.
	
A $\mu$C can be classified as programmable hardware platform that enables Embedded System applications in specific cases. It is important to know that $\mu$Cs are mainly composed of a General-Purpose Processor (GPP) of 8, 16 or 32 bits. Then this GPP is connected to some peripherals like Random Access Memory (RAM), flash memory, counters, signal generators, communication protocol specific hardware, analog to digital and digital to analog converters and others.
	
An important fact is that on most products that are available today, the $\mu$Cs embedded into them encapsulate an 8-bit GPP with enough computational power and memory storage, to show itself as a resourceful platform for many embedded applications. However, those same 8-bit $\mu$Cs are considered low-power and low-cost platforms when compared with other platforms that are used to implement AI applications with  Artificial Neural Networks (ANNs) \cite{MCUGeral1,RefG4}.
	
The use of ANNs for embedded intelligent systems with real-time constraints has been a recurrent research topic for many \cite{Ref6,Ref10,RefG1,RefG2,RefG3}. A large part of the works devised from this topic is driven by the growing demand for AI techniques for IoT, M2M and Edge Computing applications.
	
A major problem with implementing ANN applications into embedded systems is the computational complexity associated with ANNs. In regards to the Multi-Layer Perceptron (MLP), described in this work, there are many inherent multiplications and calculations of nonlinear functions \cite{Ref6,Ref10,RefG1}. Besides the feedforward process between the input and the synaptic weights, the MLP also has a training algorithm associated with it to find the optimum weights of the neural network. This training algorithm is very computationally expensive \cite{B1}. If the training process is also performed in real-time, the computational complexity is increased several times. This increase in complexity automatically raises the processing time and requirements for memory storage from the hardware platform used in the application \cite{Ref6,Ref10,RefG1}.
	
The use of MLP neural networks for real-time applications on $\mu$Cs it’s not a new effort. The work \cite{Ref7} in which the authors describe a method to linearize the nonlinear characteristics of many sensors using an MLP-NN on an 8-bit PIC18F45J10 $\mu$C. The obtained results showed that if the network architecture is right, even very difficult problems of linearization can be solved with just a few neurons.
	
In \cite{Ref3}, a fully-connected multi-layer ANN is implemented on a low-end and inexpensive $\mu$C. Also, a pseudo-floating-point multiplication is devised, to make use of the internal multiplier circuit inside the PIC\-18\-F45\-J10 $\mu$C used. The authors managed to store 256 weights into the 1KB SRAM of the $\mu$C and deemed it being enough for most embedded applications.
	
In \cite{Ref1} Farooq \textit{et al} implemented a hurdle-avoidance system controller for a car-like robot using an AT89C52 $\mu$C as a system embedding platform. They implemented an MLP with a back-propagation training algorithm and a single hidden layer. The proposed system was tested in various environments containing obstacles and was found to avoid obstacles successfully.
	
The paper \cite{saad-saoud} presents a neural network that is trained with the backpropagation (BP) algorithm and validated using a low-end and inexpensive PIC16F876A 8-bit $\mu$C. The authors chose a chemical process as a realistic example of a nonlinear system to demonstrate the feasibility and performance of this approach, as well as the results found using the microcontroller, against a computer implementation. With three inputs, five hidden neurons and an output neuron on the MLP, the application showed complete suitability for a $\mu$C-based approach. The results comparing the $\mu$C implementation showed almost no difference in Mean Square Error (MSE) after 30 iterations of the training algorithm.
	
The work presented in \cite{Ref8} an ANN-based PID controller is shown using an ARM9 based $\mu$C. The authors modeled the controller to overcome the nonlinearity of a microbiologic fermentation process and to provide a better performing control strategy. The results showed a more accurate control over the controlled parameters with an ANN-PID an achieve a greater control performance and ability of the system to meet the requirements.
	
In \cite{gural19} the authors implemented a classification application for the MNIST dataset. This implementation regards the 10 digits and full classification with $99.15\%$ testing accuracy. Also, this is implemented with a knowingly highly resource-hungry ANN model, the Convolutional Neural Network (CNN), using less than $2KB$ of SRAM memory and also $6KB$ of program memory, FLASH. This work was embedded on a Arduino Uno development kit that is comprised of a breakout board for the 8-bit ATMega$328$p $\mu$C, with $32KB$ of program memory and $2KB$ of work memory, running at $16MHz$.
	
The reference works presented above have shown different aspects of implementing ANNs on $\mu$Cs. In \cite{Ref3,Ref1,saad-saoud} you will find application proposals showing MLP-ANNs trained with the Backpropagation algorithm (BP) implemented on $\mu$Cs with good results, but none of those talk about memory usage and processing-time parameters that vary according to the MLP hyperparameters. In \cite{Ref7,Ref5,Ref8} the authors presented some results regarding processing-time, but none dependant on the number of artificial neurons or even comparing the time required to train the ANN in real-time or not.
	
Therefore, this work proposes an implementation of an MLP-ANN that can be trained with the BP algorithm into an ATMega2560 8-bit $\mu$C in the C language, to show that many applications with ANNs are suitable on this $\mu$C platform. We also present two implementations regarding a model that is trained on the $\mu$C in real-time and another implementation that is trained with Matlab and then ported to the same architecture to execute classification in real-time.
	
In addition to the implementation proposal, parameters of processing time to each feedforward step and backward steps in the training and classification process are presented. Also, the variation of these parameters is shown due to the variation of the hyperparameters of the MLP. We also validate the classification and training results on a Hardware-in-the-Loop (HIL) strategy.

\section{System Description}
	
\begin{figure*}
		\centering
		\includegraphics[width=0.7\linewidth]{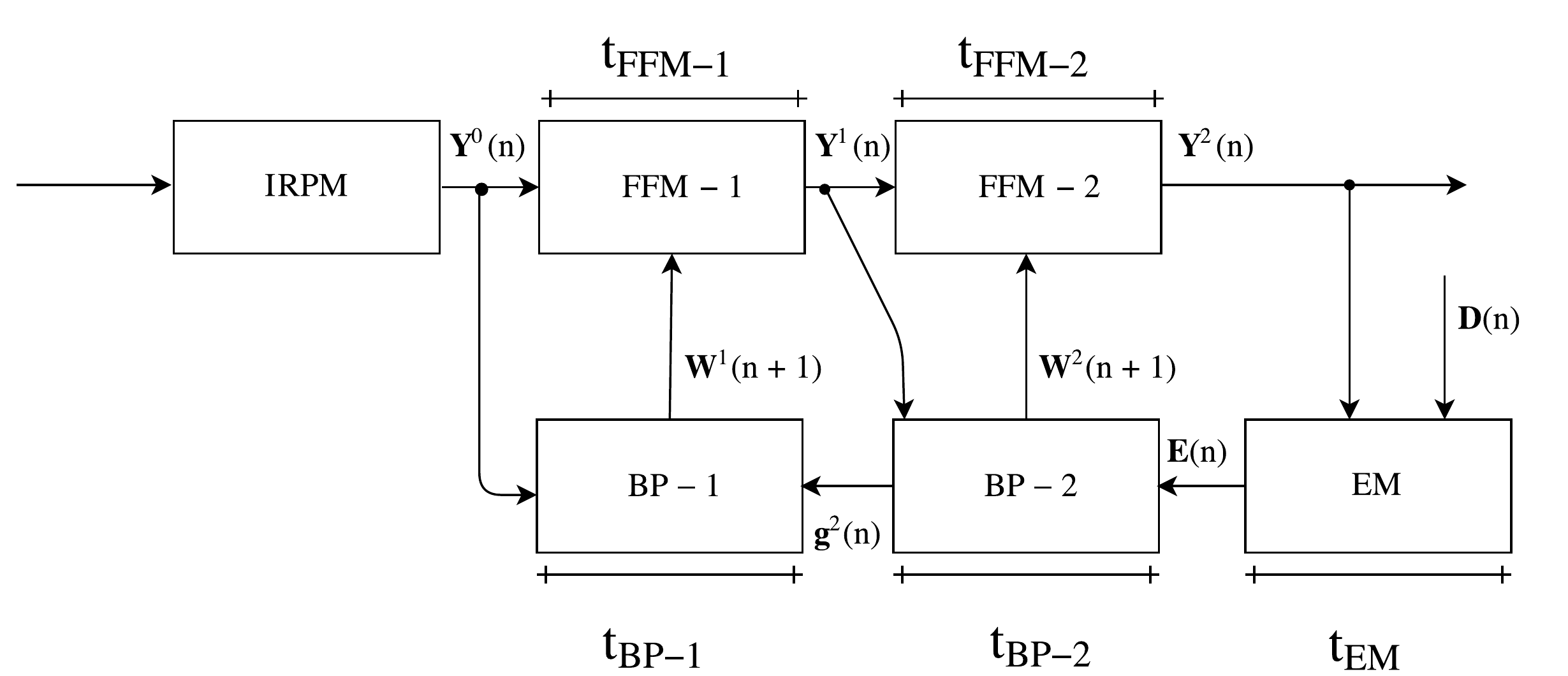}
		\caption{Block diagram detailing the modules of the matrix implementation for a MLP trained with BP algorithm.}
		\label{fig:systemdescription}
	\end{figure*}

Figure \ref{fig:systemdescription} shows a block diagram detailing the modules implemented into the $\mu$C. This implementation has the work \cite{TB1} as a direct reference, that models the MLP into a matrix form, simplifying and modularizing all the feedforward and backwards propagations As seen in Figure \ref{fig:systemdescription}, the MLP here implemented is structured as four main modules, the Input Random Permutation Module (IRPM), FeedForward Module (FFM-$k$), Error Module (EM) and Backpropagation Module (BPM-$k$), at which the variable $k$ represents the ANN layer. It is important to add that the implementation presented in this work is shown with two layers (one hidden and the output layer), but it can be easily extended for more. The modules and associated mathematical modeling will be detailed in the following subsections.
	
\subsection{Associated Variables}
	
The implementation is composed of four main variables that are passed by reference between the modules. These variables are the input signals matrix, $\mathbf{Y}^{k=0}\left( n\right)$ defined as
	\begin{equation}
	\mathbf{Y}^0\left( n\right) =
	\left[ \mathbf{y}^0_1\left(
	n\right) , \ldots ,\mathbf{y}^0_s\left( n\right)
	,\ldots ,\mathbf{y}^0_{\mathbb{N}}\left( n\right) \right]
	\end{equation}
	where $k=0$ means that this matrix is dealing with the input layer of the MLP, $n$ represents the iteration number of the training algorithm. $\mathbb{N}$ is defined as the number os samples of a training set and $\mathbf{y}_s$ the $s$-th sample defined as 
	\begin{equation}
	\mathbf{y}^0_s\left( n\right) =\left[y^0_{s1}\left( n\right) ,\ldots ,y^0_{s}\left(
	n\right), \ldots, y^0_{P}\left( n\right) \right] ^T
	\end{equation}
	where $P$ is the number of available inputs of the MLP. The synaptic weights matrix from the $k$-th layer, $\mathbf{W}^k\left( n\right)$, is defined as
	\begin{equation}
	\mathbf{W}^{k}\left( n\right) =\left[ 
	\begin{array}{ccccc}
	w_{10}^k\left( n\right) &  \cdots & w_{1h}^k\left( n\right) & \cdots & w_{1H^{k-1}}^k\left( n\right) \\ 
	\vdots & \ddots &  \vdots & \ddots & \vdots \\ 
	w_{j0}^k\left( n\right) & \cdots & w_{jh}^k\left( n\right) & \cdots & 
	w_{jH^{k-1}}^k\left( n\right) \\ 
	\vdots & \ddots &  \vdots & \ddots & \vdots \\ 
	w_{H^k0}^k\left( n\right) & \cdots & w_{H^kh}^{\left( y\right) }\left( n\right) & \cdots & 
	w_{H^kH^{k-1}}^k\left( n\right)
	\end{array}
	\right],
	\end{equation}
	where $H^k$ is the number of neurons from the $k$-th layer and $w_{ij}^k(n)$ represents the synaptic weight associated with the $i$-th artificial neuron, from the $j$-th input signal of the $k$-th layer at the $n$-th iteration. The output signals matrix, $\mathbf{Y}^L \left( n\right)$ defined as
	\begin{equation}
	\mathbf{Y}^L\left( n\right) =\left[ \mathbf{y}^L_1\left(
	n\right), \ldots ,\mathbf{y}^L_s\left( n\right)
	,\ldots ,\mathbf{y}^L_{\mathbb{N}}\left( n\right) \right],
	\end{equation}
	where $L$ is the number of layers of the MLP and also defines which layer is the last one. Where $y_s^L(n)$ represents the output signal associated with the $s$-th sample input $y_s^0$ that is defined as
	\begin{equation}
	\mathbf{y}^L_s\left( n\right) =\left[ y^L_{1}\left( n\right) ,\ldots,y^L_{s}\left(
	n\right) ,\ldots,y^L_{M}\left( n\right) \right] ^T
	\end{equation}
	where $M$ is the number of output neurons.
	
The $\mathbf{D}\left( n\right)$ variable represents the desired values or labels of the training set composed of the $\mathbf{D}\left( n\right)$ and the $\mathbf{Y}^0\left( n\right)$, that is defined as
	\begin{equation}
	\mathbf{D}\left( n\right) =\left[ \mathbf{d}_1\left(
	n\right) ,\mathbf{d}_2\left( n\right) ,\ldots ,\mathbf{d}_s\left( n\right)
	,\ldots ,\mathbf{d}_{\mathbb{N}}\left( n\right) \right]
	\end{equation}
where $\mathbf{d}_s\left( n\right)$ is the vector of desired values referring to the $s$-th sample associated with the $y_s^0(n)$ input signal, that is defined as
	\begin{equation}
	\mathbf{d}_s\left( n\right) =\left[ d_{1}\left( n\right) ,\ldots ,d_{s}\left( n\right),\ldots,d_{M}\left( n\right) \right] ^T.
	\end{equation}
	
For an MLP with two layers,  $\mathbf{W}^1\left( n\right)$ represents the synaptic weights matrix from the hidden layer ($H^0 = P$) and $\mathbf{W}^2\left( n\right)$ is the weights matrix from the output layer ($H^2 = M$) at the $n$-th iteration. In addition to those two matrices, a few others are created to accommodate intermediary results from feedforward and backward propagation operations of the MLP.
	
\subsection{\emph{Feedforward Module} - ($\text{FFM}-k$)}
	
This module is responsible for running the feedforward operation of an MLP, propagating the inputs through each $k$-th layer during each $n$-th iteration. As most BP implementations, this proposal can operate in online mode or batch mode (defining $\mathbb{N} > 1$). At each $k$-th FFM-$k$ at each $k$-th layer the following equation is calculated as
	\begin{equation}
	\mathbf{Y}^k \left( n\right) = \varphi \left( \mathbf{W}^k\left( n\right)  \mathbf{Y}^{k-1}\left( n\right)\right)
	\label{eqn:feedforward}
	\end{equation} 
	where $\mathbf{Y}^{k-1}\left( n\right)$ is the output signals matrix from the previous layer. The $\mathbf{Y}^{k}\left( n\right)$ represents the output of the current layer and $\varphi \left(.\right)$ is the activation function of the current $k$-th layer. As previously said a few other matrices are devised as part of the BP calculation and some of them store these calculation results from Equation \ref{eqn:feedforward}.
	
	Algorithm \ref{AFFM} describes the pseudocode executed when the FFM-$k$ module is called. The function \textit{prodMatrix()} implements the matrix product between $\mathbf{W}^k\left( n\right)$ and $\mathbf{Y}^{k-1}\left( n\right)$ and in the end it stores the product result in $\mathbf{Y}^{k}\left( n\right)$. The \textit{actFun()} function executes the activation function of the $k$-th layer to each element of $\mathbf{Y}^{k}\left( n\right)$ and stores the result in itself ($\mathbf{Y}^{k}\left( n\right)$) as it runs.
	
	\begin{algorithm}[ht]
		\caption{Description of the algorithm implemented by the $\text{FFM}-k$ module.}
		\label{AFFM}	
		\begin{algorithmic}[1]
			\setstretch{2.0}
			\Function{$\text{FFM}-k$}{$\mathbf{W}^k\left( n\right)$,$\mathbf{Y}^{k-1}\left( n\right)$,$\mathbf{Y}^{k}\left( n\right)$}			
			\State $\text{prodMat}\left(\mathbf{W}^k\left( n\right),\mathbf{Y}^{k-1}\left( n\right),\mathbf{Y}^{k}\left( n\right)\right)$			
			\State $\text{activfun}\left(\mathbf{Y}^{k}\left( n\right) \right)$			
			\EndFunction
		\end{algorithmic}
	\end{algorithm}
	
	\subsection{\emph{Error Module} - ($\text{EM}-k$)}
	The error module EM-$k$ calculates the error between the desired values matrix $\mathbf{D}\left( n\right)$ and the output signals matrix $\mathbf{Y}^L\left( n\right)$ of the last layer, $L$. The equation that is evaluated at the EM-$k$ is defined as
	\begin{equation}
	\mathbf{E}\left( n\right) = \mathbf{D}\left( n\right)  - \mathbf{Y}^L\left( n\right) 
	\end{equation}
	
	Algorithm \ref{AEM} describes how the EM-$k$ module functions, in which the \textit{difMatrix()} function implements the element-by-element difference between $\mathbf{D}\left( n\right)$ and $\mathbf{Y}^L\left( n\right)$ and stores the result into the $\mathbf{E}\left( n\right)$ matrix.
	
	\begin{algorithm}[ht]
		\caption{Pseudocode implementation of the  $\text{EM}$ module.}
		\label{AEM}
		\begin{algorithmic}[1]
			\setstretch{2.0}
			\Function{$\text{EM}$}{$\mathbf{Y}^L\left( n\right)$,$\mathbf{D}\left( n\right)$,$\mathbf{E}\left( n\right)$}
			\State $\text{difMatrix}\left(\mathbf{Y}^{L}\left( n\right),\mathbf{D}\left( n\right), \mathbf{E}\left( n\right)\right)$
			\EndFunction
		\end{algorithmic}
	\end{algorithm}
	
	\subsection{\emph{Backpropagation Module} - ($\text{BPM}-k$)}
	
	The last module is responsible for the main training part of the implementation. The BPM-$k$ calculates the new updated values of the synaptic weights matrices of the $L$ layers that better approximate the desired values, $\mathbf{W}^k\left( n\right)$. The equation that the BPM-$k$ implements is defined as
	\begin{equation}
	\mathbf{W}^{\left( k\right) }\left( n+1\right) =\mathbf{W}^{\left(k\right)}\left( n\right) +\Delta \mathbf{W}^{\left( k\right) }\left( n\right) 
	\end{equation}
	where \begin{equation}
	\Delta \mathbf{W}^{\left( k\right) }\left( n\right)  = \frac{\eta}{\mathbb{N}}\mathbf{g}^k\left( n\right) \left[ 
	\begin{array}{c}
	\mathbf{-1} \\ 
	\mathbf{Y}^{k-1}\left( n\right)
	\end{array}
	\right] ^T + \alpha \mathbf{\Delta W}^{\left( k\right) }\left( n-1\right) ,
	\end{equation}
	where $\eta$ is the learning-rate of the BP algorithm, $\alpha$ is the learning-moment rate factor and $\mathbf{g}^k\left( n\right)$ is defined as
	\begin{equation}
	\mathbf{g}^k\left( n\right) = \left \{
	\begin{array}{ll}
	\text{prod}\left( \varphi ^{\prime
	}\left( \mathbf{Y}^k\left( n\right) \right) ,\mathbf{E}\left( n\right) \right) & \text{ for } k=L \\
	\mathbf{z}^k\left( n\right) & \text{ for } 1 \leq k <  L
	\end{array}
	\right.
	\end{equation}
	where 
	\begin{equation}
	\mathbf{z}^k\left( n\right) = \text{prod}\left(
	\varphi ^{\prime }\left( \left[ 
	\begin{array}{c}
	\mathbf{-1} \\ 
	\mathbf{Y}^k\left( n\right)
	\end{array}
	\right] \right) ,\left[ \mathbf{W}^{\left( y\right) }\left( n\right) \right]
	^T \mathbf{g}^{\left( k+1\right) }\left( n\right)  \right) 
	\label{eqn:prodab}
	\end{equation}
	and
	\begin{equation}
	\mathbf{z}^k\left( n\right)  = \left[ 
	\begin{array}{c}
	\begin{array}{cccc}
	z_{11}^{\left( k\right) }\left( n\right) & z _{12}^{\left(
		k\right) }\left( n\right) & \ldots & z _{1\mathbb{N}}^{\left( k\right)
	}\left( n\right)
	\end{array}
	\\ \hline
	\mathbf{z}^{\left( k\right) }\left( n\right)
	\end{array}
	\right].
	\end{equation}
	
	The \textit{prod()} function in the Equation \ref{eqn:prodab} implements an element-wise product between two matrices, implemented in Algorithm \ref{PRODMAT}, and $\varphi ^{\prime} (\cdot)$ is the derivative of the activation function.

	\subsection{Basic Operations}
	
	In this section, we describe how we implemented a few basic operations used in this work. First, Algorithm \ref{PRODMAT} implements the matrix product. A very important detail is that all the modules were implemented following pointer arithmetic procedures. With this in mind, the source code is very optimized for memory usage, reducing the overall duration of execution from each module. Every matrix used in this implementation are floating-point (IEEE754) \cite{ieee754} representation values.
	
	\begin{algorithm}[ht]
		\caption{Matrix product implementation pseudocode}
		\label{PRODMAT}		
		\begin{algorithmic}[1]
			\setstretch{2.0}
			\Function{$prodMat$}{$\mathbf{W^k}, \mathbf{Y^{k-1}}, \mathbf{Y^k}, P, M, \mathbb{N}$}			
			\State{Initialize ($i$, $j$, $s$, accumulate) $\leftarrow$ 0}			
			\For{$i$ $\leftarrow$ 1 to M}			
			\For{$j$ $\leftarrow$ 1 to $\mathbb{N}$}			
			\State{accumulate $\leftarrow$ 0}
			\For{$s$ $\leftarrow$ 1 to P}
			\State{accumulate = accumulate + $\mathbf{W^k}$[$i$][$s$]$\times$ $\mathbf{Y^{k-1}}$[$s$][$j$]}
			\EndFor
			\State{$\mathbf{Y^k}$[$i$][$j$] = accumulate}
			\EndFor
			\EndFor
			\EndFunction		
		\end{algorithmic}		
	\end{algorithm}
	
	Algorithm \ref{PRODPTT} shows the pseucode of a element-wise product between two matrices used in this work, more specifically in the BPM-$k$ module implementation. This same pseudocode in Algorithm \ref{PRODPTT} can be slightly altered to perform subtraction or addition by editing the operation between matrices on line $5$. It's important to notice that Algorithm \ref{PRODPTT} requires that both matrices have the same dimensions.
	
	\begin{algorithm}[ht]
		\caption{Element-wise matrix product implementation algorithm.
			\label{PRODPTT}}		
		\begin{algorithmic}[1]
			\setstretch{2.0}
			\Function{$prod$}{$\mathbf{Y^k}, \mathbf{E}, \mathbf{g^k}, P,M$}
			\State{Initialise ($i$, $j$) $\leftarrow$ 0}
			\For{$i$ $\leftarrow$ 1 to P}
			\For{$j$ $\leftarrow$ 1 to M}
			\State{$\mathbf{g^k}$[$i$][$j$] = $\mathbf{Y^k}$[$i$][$j$]$\times$$\mathbf{E}$[$i$][$j$]}	
			\EndFor
			\EndFor
			\EndFunction
		\end{algorithmic}		
	\end{algorithm}
	
	Algorithm \ref{PRODTRACE} presents to us the implementation of the calculation of the \textit{Trace} and \textit{Product} between two matrices at a single operation inside a nested loop. This simple optimization saved a few $\mu$s of processing time for the overall training process.
	
	\begin{algorithm}[ht]
		\caption{Pseudocode implementation of the trace of the product between two matrices. \label{PRODTRACE}}		
		\begin{algorithmic}[1]
			\setstretch{2.0}
			\Function{$trace$}{$\mathbf{E^k}, \mathbf{E^{kT}}, P, M$}
			\State{Initialize ($i$, $j$,trace) $\leftarrow$ 0}
			\For{$i$ $\leftarrow$ 1 to P}
			\For{$j$ $\leftarrow$ 1 to M}
			\State{trace = trace +  $\mathbf{E^k}$[$i$][$j$]$\times$ $\mathbf{E^kT}$[$i$][$j$]}	
			\EndFor
			\EndFor
			\EndFunction		
		\end{algorithmic}		
	\end{algorithm}
	
	Algorithm \ref{activFun} shows the implemented steps for calculating the activation function of the induced-local-field of each neuron, an element-wise operation. This Algorithm \ref{activFun} is implementing a sigmoid activation function.
	
	\begin{algorithm}[ht]
		\caption{Pseudocode implementation of the sigmoid activation function \label{activFun}}		
		\begin{algorithmic}[1]
			\setstretch{2.0}
			\Function{$activfun$}{$\mathbf{Y^k}, \mathbf{Y^{k+1}}, P, M$}
			\State{Initialise ($i$, $j$) $\leftarrow$ 0}
			\For{$i$ $\leftarrow$ 1 to P}
			\For{$j$ $\leftarrow$ 1 to M}
			\State{$\mathbf{y^{k+1}}$[$i$][$j$]=(1.0)/(1.0+exponential(-$\mathbf{Y^k}$[$i$][$j$]))}	
			\EndFor
			\EndFor
			\EndFunction		
		\end{algorithmic}		
	\end{algorithm}
	
	\section{Methodology}
	
	The implementation was validated using the HIL simulation strategy, as explained in \cite{RefG4} and illustrated in Figure \ref{fig:hilxor}. The tested variables are, execution time for the FFM-$1$ for the hidden layer as seen in Figure \ref{fig:systemdescription} defined as $t_{\text{FFM-}1}$, execution time for the FFM-$2$, feedforward for the output layer of neurons,$t_{\text{FFM-}2}$, error module $t_{\text{EM}}$, backpropagation of the output neurons layer, $t_{\text{BPM-}1}$, and the backpropagation of the hidden neurons layer, $t_{\text{BPM-}2}$.
	
	A parameter that was also analyzed is the memory occupation, regarding both memories of the ATMega-$2560$ $\mu$C used in this paper, FLASH or program memory and SRAM or work memory.
	
	As previously mentioned the modules were implemented in the C program language for AVR $\mu$Cs using the \textit{avr-gcc} version 5.4.0, inside the Atmel Studio 7 development environment, an Integrated Development Environment (IDE) made available by Microchip. After compilation and binary code generation the solution was embedded into an Atmega-$2560$. This $\mu$C has an 8-bit GPP integrated with 256KBytes of FLASH program memory and 8KBytes of SRAM work memory, its maximum processing speed is 1 MIPS/MHz.
	
	The Atmega-$2560$ is associated to an Arduino Mega v2.0 development kit, the Arduino Mega is a development kit that provides a breakout board for all the ATmega-$2560$ pins and some other components required for the $\mu$C to function properly. A great feature of this development kit is the onboard Universal Serial Bus (USB) programmer that enables the developer to simply connect a USB port to a computer and test various implementations easily on the $\mu$C.
	
	This work is further validated using two cases. First, we train the MLP-BP to behave as an XOR operation as the simplest case possible to train an MLP and evaluate its ability to learn a nonlinear relationship between two inputs. Secondly, we train the network to aid a car-like virtual robot in avoiding obstacles in a virtual map using Matlab and 3 cases of increasing ANN architecture complexity. The assembly, and analysis of these two validation cases are presented in the following subsections.
	
	\subsection{Hardware in the Loop Simulation}
	
\begin{figure*}
		\centering
		\includegraphics[width=0.6\linewidth]{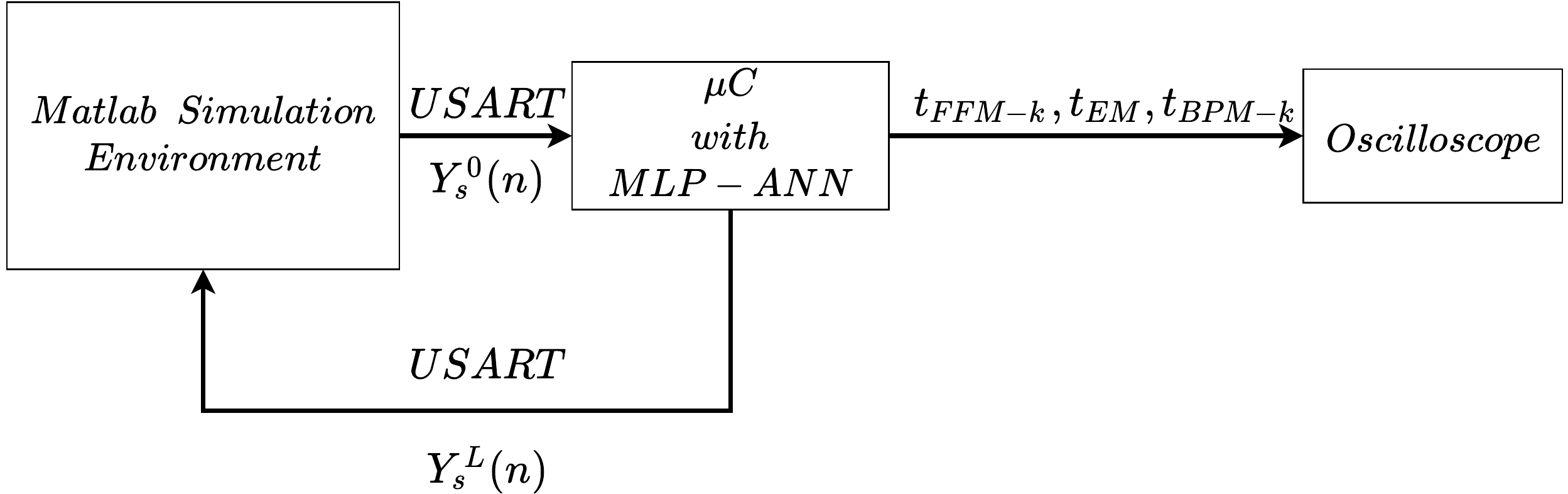}
		\caption{Hardware In the Loop assembly scheme in block diagram.}
		\label{fig:hilxor}
	\end{figure*}

\begin{figure}
		\centering
		\includegraphics[width=0.9\linewidth]{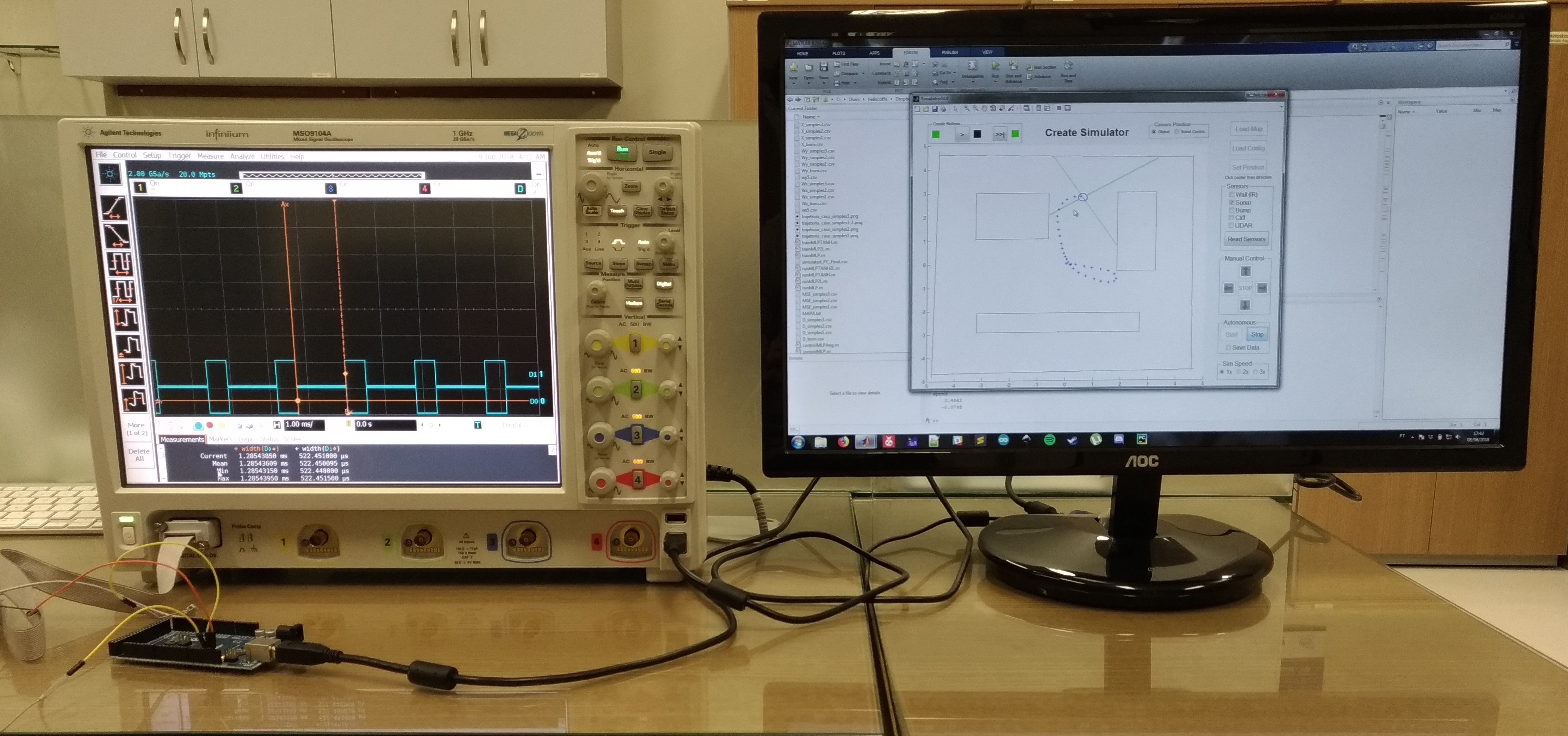}
		\caption{Hardware In the Loop assembly on the test bench.}
		\label{fig:fotomontagemhil}
	\end{figure}

\begin{figure}
		\centering
		\includegraphics[width=0.8\linewidth]{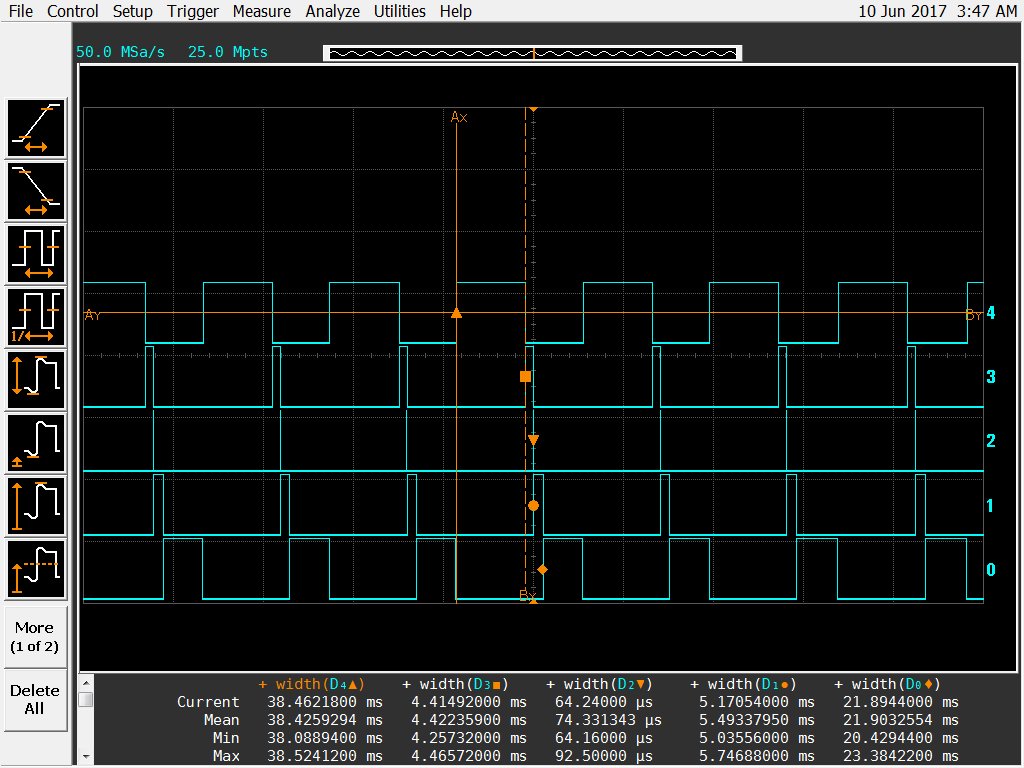}
		\caption{Hardware In the Loop measurements of the embedded XOR operation using an oscilloscope.}
		\label{fig:38-neuronsalldata}
	\end{figure}

	The tests were executed with the $\mu$C running at a clock of $16$MHz. The results are obtained by setting the level of a digital pin of the ATMega-$2560$ to HIGH, executing one of the modules and setting the same digital pin logical value to LOW and measuring the time of logic HIGH on an oscilloscope for this digital pin. This can be easily seen on Figure \ref{fig:hilxor} and a picture of the HIL assembly is shown in Figure \ref{fig:fotomontagemhil}. In Figures \ref{FigResults12},\ref{FigResults34} we present curved plots of the results described into \ref{TabResults1} where we performed curve-fitting on the measured points with Polynomial Regression. Figure \ref{fig:38-neuronsalldata} shows a closer look on the oscilloscope measurements, taking in regard all the modules being measured by their duration of execution for the XOR validation case described below.

	The calculation of the MSE during training was also embedded into the $\mu$C implementation and they're transmitted through Universal Serial Synchronous Asynchronous Receiver Transmitter (USART) protocol to a computer. The calculation of the MSE is defined as
	\begin{equation}
	MSE(n) = \frac 1{2N}\text{trace}\left(\mathbf{E}(n)^T\mathbf{E}(n)\right)
	\end{equation}
	where $\text{trace}(\cdot)$ is the implementation seen in Algorithm \ref{PRODTRACE}.
	
	\subsection{XOR Operation}
	
A validation of the real-time embedded training using the $\text{BPM}-k$ module was devised for a XOR operation, as described on Table \ref{tab:xor}, the most basic MLP validation case. This test was performed with a training and classification phases, both running in the $\mu$C, considering a configuration as seen in Figure \ref{fig:systemdescription}. The test was executed on batch mode with $\mathbb{N} = 4$, two layers of neurons ($L = 2$), two input signals ($P = H^0 = 2$), a varying amount of neurons in the hidden neurons layer ($H^1$) as seen in \ref{TabResults1} and a single neuron in the output layer ($H^2 = 1$). It is important to notice that the strategy here presented can be assembled with various different configurations just modifying the $P,H^k$ and $M$ parameters, being the $\mu$Cs internal memories the limiting factor for the ANN architecture size.

	\begin{table}[ht]
		\centering
		\begin{tabular}{|c|c|c|}
			\hline
			Input A & Input B & Result \\ \hline
			0    &    0    &   0    \\ \hline
			0    &    1    &   1    \\ \hline
			1    &    0    &   1    \\ \hline
			1    &    1    &   0    \\ \hline
		\end{tabular}
		\caption{Truth table for the XOR operation}
		\label{tab:xor}
	\end{table}

\subsection{Virtual Car-Like Robot}
	
	This work also tested a MLP-BP model to control a virtual car-like robot from the \emph{iRobot} matlab toolbox, provided by the United States Naval Academy (USNA), \cite{irobotcreate}. This toolbox provides a virtual environment and an interface to test control algorithms on a deferentially steered robot on various maps with different obstacles and different combinations of distance or proximity sensors.
	
	As previously stated, this virtual environment provides an interface to control a deferentially steered virtual robot. This robot is controlled by changing the angular speed in $rad/s$ of the two wheels, Right Wheel (RW) and Left Wheel (LW). Also this work used three proximity sensors with virtual 3 meters range to provide input to the MLP. The sensors are Front-Sensor (FS), Left-Sensor (LS) and Right-Sensor (RS).
	
	We tested the MLP model with three different datasets, increasing the hyperparameters to evaluate the behavior of the robot as we increased the network architecture's complexity. The cases are comprised of three tables with five columns each, as seen in Table \ref{tab:casosimples1}. The first case is very simple with only eight conditions provided to train the network, which means the network must be able to devise a knowledge representation on how to behave with cases not previously trained from these simple constraints. 
	
	\begin{table*}[ht]
		\begin{center}
			\begin{tabular}{|c|c|c|c|c|c|}
				\hline
				Condition &   FS (m)    &   RS (m)    &   LS (m)    &  LW (rad/s)  &  RW (rad/s)  \\ \hline
				1     & 0\text{.}00 & 0\text{.}00 & 0\text{.}00 & 0\text{.}50  & -0\text{.}50 \\ \hline
				2     & 0\text{.}00 & 0\text{.}00 & 1\text{.}00 & -0\text{.}50 & 0\text{.}50  \\ \hline
				3     & 0\text{.}00 & 1\text{.}00 & 0\text{.}00 & 0\text{.}50  & -0\text{.}50 \\ \hline
				4     & 0\text{.}00 & 1\text{.}00 & 1\text{.}00 & 0\text{.}50  & -0\text{.}50 \\ \hline
				5     & 1\text{.}00 & 0\text{.}00 & 0\text{.}00 & 0\text{.}50  & 0\text{.}50  \\ \hline
				6     & 1\text{.}00 & 0\text{.}00 & 1\text{.}00 & 0\text{.}50  & 0\text{.}50  \\ \hline
				7     & 1\text{.}00 & 1\text{.}00 & 0\text{.}00 & 0\text{.}50  & 0\text{.}50  \\ \hline
				8     & 1\text{.}00 & 1\text{.}00 & 1\text{.}00 & 0\text{.}50  & 0\text{.}50  \\ \hline
			\end{tabular}
		\end{center}
		\caption{Training dataset for the first and simplest case. With $H^0 = 3$, $H^1 = 5$ and $H^2 = 2$.}
		\label{tab:casosimples1}
	\end{table*}
	
The second case has a greater complexity with 18 conditions, as seen in Table \ref{tab:casosimples2}. Also, as mentioned in Table \ref{tab:casosimples2} this case required the same amount of neurons in the hidden layer, meaning that the first architecture could still be used for a more complex case.
	
	\begin{table*}[ht]
		\begin{center}
			\begin{tabular}{|c|c|c|c|c|c|}
				\hline
				Condition &   FS (m)    &   RS (m)    &    LS (m)    &  LW (rad/s)  &  RW (rad/s)  \\ \hline
				1     & 0\text{.}00 & 0\text{.}00 & 0\text{.}00  & -0\text{.}20 & 0\text{.}20  \\ \hline
				2     & 1\text{.}00 & 0\text{.}00 & 0\text{.}00  & -0\text{.}20 & 0\text{.}20  \\ \hline
				3     & 2\text{.}00 & 0\text{.}00 & 1\text{.}00  & -0\text{.}20 & 0\text{.}20  \\ \hline
				4     & 0\text{.}00 & 1\text{.}00 & 1\text{.}00  & 0\text{.}10  & 0\text{.}10  \\ \hline
				5     & 1\text{.}00 & 1\text{.}00 & 2\text{.}00  & 0\text{.}20  & -0\text{.}10 \\ \hline
				6     & 2\text{.}00 & 1\text{.}00 & 2\text{.}00  & 0\text{.}20  & -0\text{.}10 \\ \hline
				7     & 0\text{.}00 & 2\text{.}00 & 0\text{.}00  & 0\text{.}50  & 0\text{.}50  \\ \hline
				8     & 1\text{.}00 & 2\text{.}00 & 0\text{.}00  & 0\text{.}50  & 0\text{.}50  \\ \hline
				9     & 2\text{.}00 & 2\text{.}00 & 1\text{.}00  & 0\text{.}50  & 0\text{.}50  \\ \hline
				10     & 0\text{.}00 & 0\text{.}00 & 1\text{.}00  & 0\text{.}20  & 0\text{.}20  \\ \hline
				11     & 1\text{.}00 & 0\text{.}00 & 2\text{.}00  & 0\text{.}20  & -0\text{.}20 \\ \hline
				12     & 2\text{.}00 & 0\text{.}00 & 2\text{.}00  & 0\text{.}20  & -0\text{.}20 \\ \hline
				13     & 0\text{.}00 & 1\text{.}00 & 0\text{.}00  & 0\text{.}10  & 0\text{.}10  \\ \hline
				14     & 1\text{.}00 & 1\text{.}00 & 0\text{.}00  & 0\text{.}10  & 0\text{.}10  \\ \hline
				15     & 2\text{.}00 & 1\text{.}00 & 1\text{.}00  & -0\text{.}10 & 0\text{.}10  \\ \hline
				16     & 0\text{.}00 & 2\text{.}00 & 10\text{.}00 & 0\text{.}50  & 0\text{.}50  \\ \hline
				17     & 1\text{.}00 & 2\text{.}00 & 2\text{.}00  & 0\text{.}50  & 0\text{.}50  \\ \hline
				18     & 2\text{.}00 & 2\text{.}00 & 2\text{.}00  & 0\text{.}50  & 0\text{.}50  \\ \hline
			\end{tabular}
		\end{center}
		\caption{Training dataset for the second case. With $H^0 = 3$, $H^1 = 5$ and $H^2 = 2$.}
		\label{tab:casosimples2}
	\end{table*}
	
The third and most complex case with 27 conditions to train the MLP-BP. This dataset required a bigger architecture than the previous datasets, with double the amount of neurons in the hidden layer ($H^1 = 10$)
	
	\begin{table*}[ht]
		\begin{center}
			\begin{tabular}{|c|c|c|c|c|c|}
				\hline
				Condition &   FS (m)    &   RS (m)    &   LS (m)    &  LW (rad/s)  &  RW (rad/s)  \\ \hline
				1     & 0\text{.}50 & 0\text{.}50 & 0\text{.}50 & 0\text{.}30  & 0\text{.}30  \\ \hline
				2     & 0\text{.}50 & 0\text{.}50 & 1\text{.}00 & 0\text{.}20  & -0\text{.}20 \\ \hline
				3     & 0\text{.}50 & 0\text{.}50 & 2\text{.}50 & 0\text{.}20  & -0\text{.}20 \\ \hline
				4     & 0\text{.}50 & 1\text{.}00 & 0\text{.}50 & 0\text{.}30  & 0\text{.}30  \\ \hline
				5     & 0\text{.}50 & 1\text{.}00 & 1\text{.}00 & 0\text{.}20  & -0\text{.}20 \\ \hline
				6     & 0\text{.}50 & 1\text{.}00 & 2\text{.}50 & 0\text{.}20  & -0\text{.}20 \\ \hline
				7     & 0\text{.}50 & 2\text{.}50 & 0\text{.}50 & 0\text{.}30  & 0\text{.}30  \\ \hline
				8     & 0\text{.}50 & 2\text{.}50 & 1\text{.}00 & 0\text{.}30  & 0\text{.}30  \\ \hline
				9     & 0\text{.}50 & 2\text{.}50 & 2\text{.}50 & 0\text{.}30  & 0\text{.}30  \\ \hline
				10     & 1\text{.}00 & 0\text{.}50 & 0\text{.}50 & -0\text{.}20 & 0\text{.}20  \\ \hline
				11     & 1\text{.}00 & 0\text{.}50 & 2\text{.}50 & 0\text{.}20  & -0\text{.}20 \\ \hline
				12     & 1\text{.}00 & 1\text{.}00 & 0\text{.}50 & -0\text{.}20 & 0\text{.}20  \\ \hline
				13     & 0\text{.}50 & 1\text{.}00 & 0\text{.}50 & 0\text{.}30  & -0\text{.}30 \\ \hline
				14     & 1\text{.}00 & 1\text{.}00 & 1\text{.}00 & 0\text{.}30  & 0\text{.}30  \\ \hline
				15     & 1\text{.}00 & 1\text{.}00 & 2\text{.}50 & 0\text{.}20  & -0\text{.}20 \\ \hline
				16     & 1\text{.}00 & 2\text{.}50 & 0\text{.}50 & -0\text{.}20 & 0\text{.}20  \\ \hline
				17     & 1\text{.}00 & 2\text{.}50 & 1\text{.}00 & 0\text{.}30  & 0\text{.}30  \\ \hline
				18     & 1\text{.}00 & 2\text{.}50 & 2\text{.}50 & 0\text{.}20  & -0\text{.}20 \\ \hline
				19     & 2\text{.}50 & 0\text{.}50 & 0\text{.}50 & -0\text{.}20 & 0\text{.}20  \\ \hline
				20     & 2\text{.}50 & 0\text{.}50 & 1\text{.}00 & -0\text{.}20 & 0\text{.}20  \\ \hline
				21     & 2\text{.}50 & 0\text{.}50 & 2\text{.}50 & -0\text{.}20 & 0\text{.}20  \\ \hline
				22     & 2\text{.}50 & 1\text{.}00 & 0\text{.}50 & -0\text{.}20 & 0\text{.}20  \\ \hline
				23     & 2\text{.}50 & 1\text{.}00 & 1\text{.}00 & -0\text{.}20 & 0\text{.}20  \\ \hline
				24     & 2\text{.}50 & 1\text{.}00 & 2\text{.}50 & 0\text{.}20  & -0\text{.}20 \\ \hline
				25     & 2\text{.}50 & 2\text{.}50 & 0\text{.}50 & -0\text{.}20 & 0\text{.}20  \\ \hline
				26     & 2\text{.}50 & 2\text{.}50 & 1\text{.}00 & -0\text{.}20 & 0\text{.}20  \\ \hline
				27     & 2\text{.}50 & 2\text{.}50 & 2\text{.}50 & 0\text{.}30  & 0\text{.}30  \\ \hline
			\end{tabular}
		\end{center}
		\caption{Training dataset for the third and most complex case. With  $H^0 = 3$, $H^1 = 10$ e $H^2 = 2$.}
		\label{tab:casosimples3}
	\end{table*}
	
	\section{Results}
	
	\subsection{XOR Operation}
	The binary code for the XOR operation, that is generated from this implementation as seen in Figure \ref{fig:systemdescription}, resulted in $6.672$ KBytes of program memory occupation (equivalent to $2.6\%$), a pretty compact solution if compared to the maximum program memory available for the ATMega-$2560$ and also comparing to the $5.904$KBytes presented in \cite{Ref4} for and MLP implementation without the BP training, considering that these $2.6\%$ already include the training algorithm.
	
	The information presented at Table \ref{TabResults1} show that the values obtained up to $H^1 = 38$ neurons in the hidden layer. The $H^1$ parameter is limited by the available memory in the $\mu$C used, ATMega-$2560$. However, this value is quite reasonable for most real-time applications in robotics and industrial automation. Another noticeable result is that the processing times are also reasonable. Analyzing the network without the training time results (EM, BPM-$1$ and BPM-$2$) we can see that the iteration for a batch mode of $\mathbb{N} = 4$ samples takes $42.91 \text{ms}$ for the worst case, $H^1 = 38$. If you analyze the duration of each iteration including the network being trained on the online mode it takes $69.88 \text{ms}$, again in the worst case of $H^1=38$. This shows that the implementation here presented is indeed suitable for commercial applications in fields like industrial automation, robotics, automotive industry, and others.
	
\begin{figure}[ht]
		\begin{center}
			\subfigure[]
			{
				\includegraphics[width=.52\textwidth]{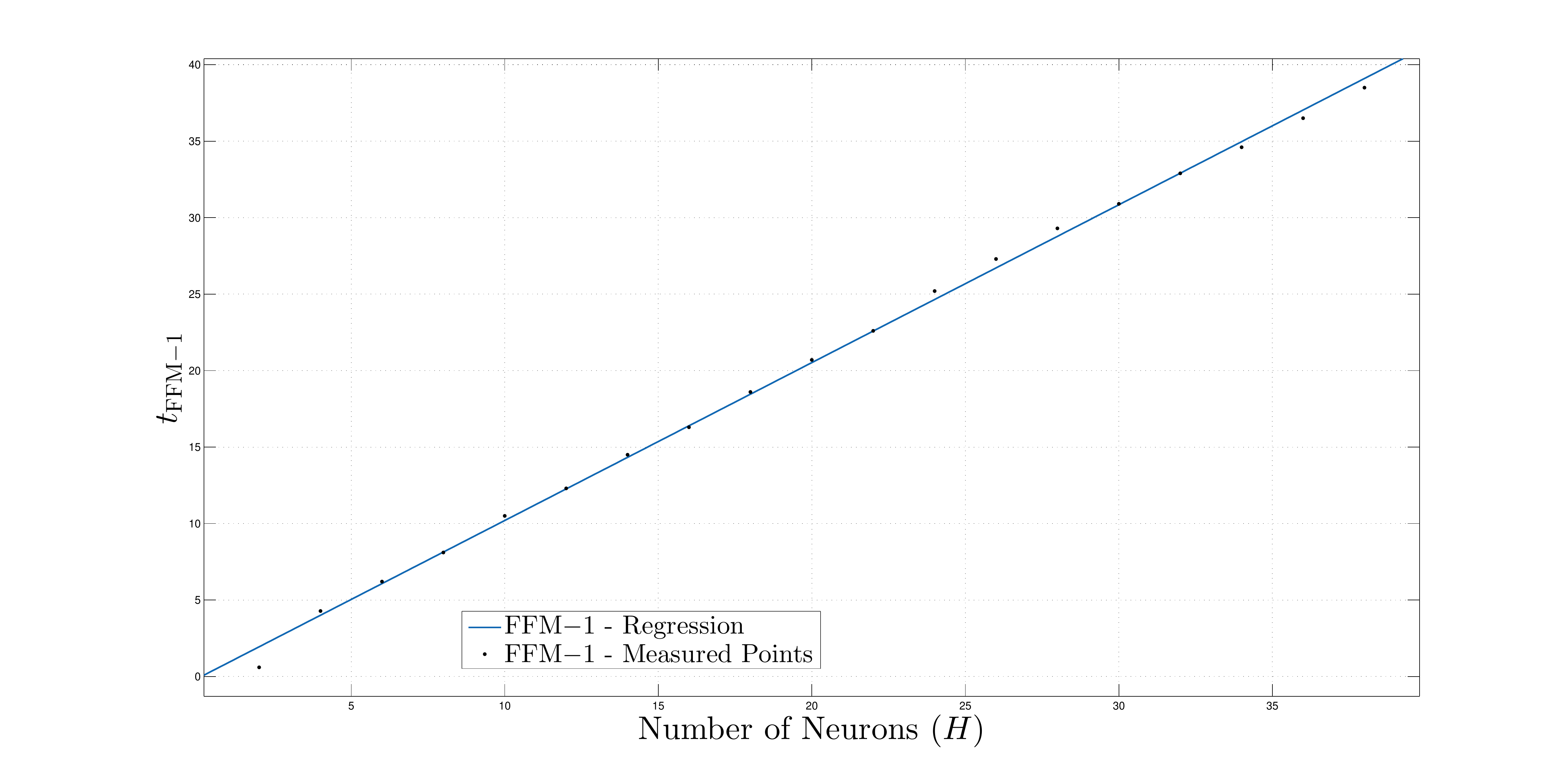}
				\label{FigResults1}
			}
			\subfigure[]
			{
				\includegraphics[width=.52\textwidth]{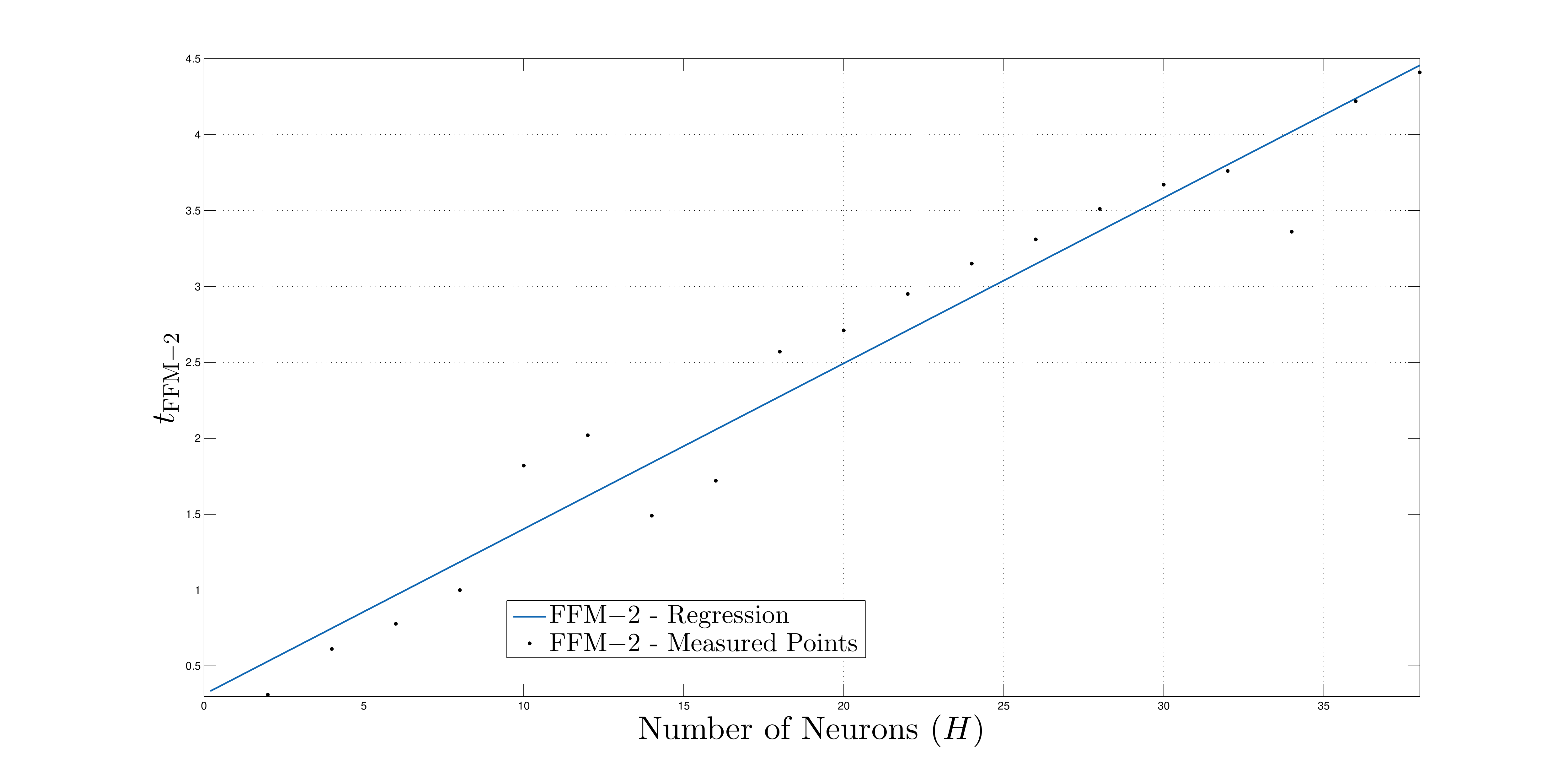}
				\label{FigResults2}
			}
			\caption{a) Execution time of $\text{FFM}-1$ by number of neurons in the hidden layer, $H^1$. b) Execution time of $\text{FFM}-2$ by the number of neurons in the hidden layer, $H^1$.}
			\label{FigResults12}
		\end{center}
	\end{figure}

\begin{figure}[ht]
		\begin{center}
			\subfigure[] {
				\includegraphics[width=.52\textwidth]{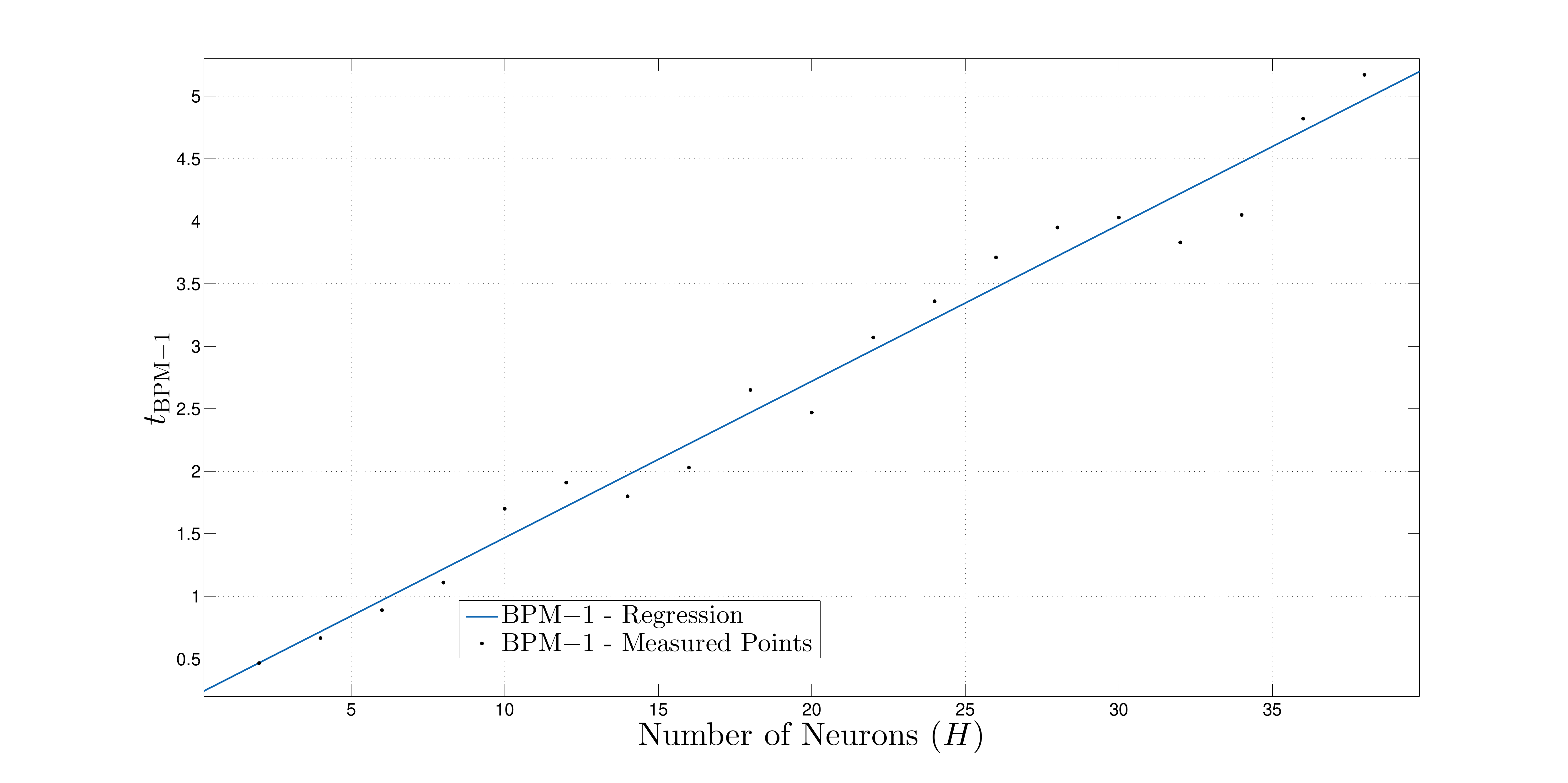}
				\label{FigResults3}
			}
			\subfigure[] {
				\includegraphics[width=.52\textwidth]{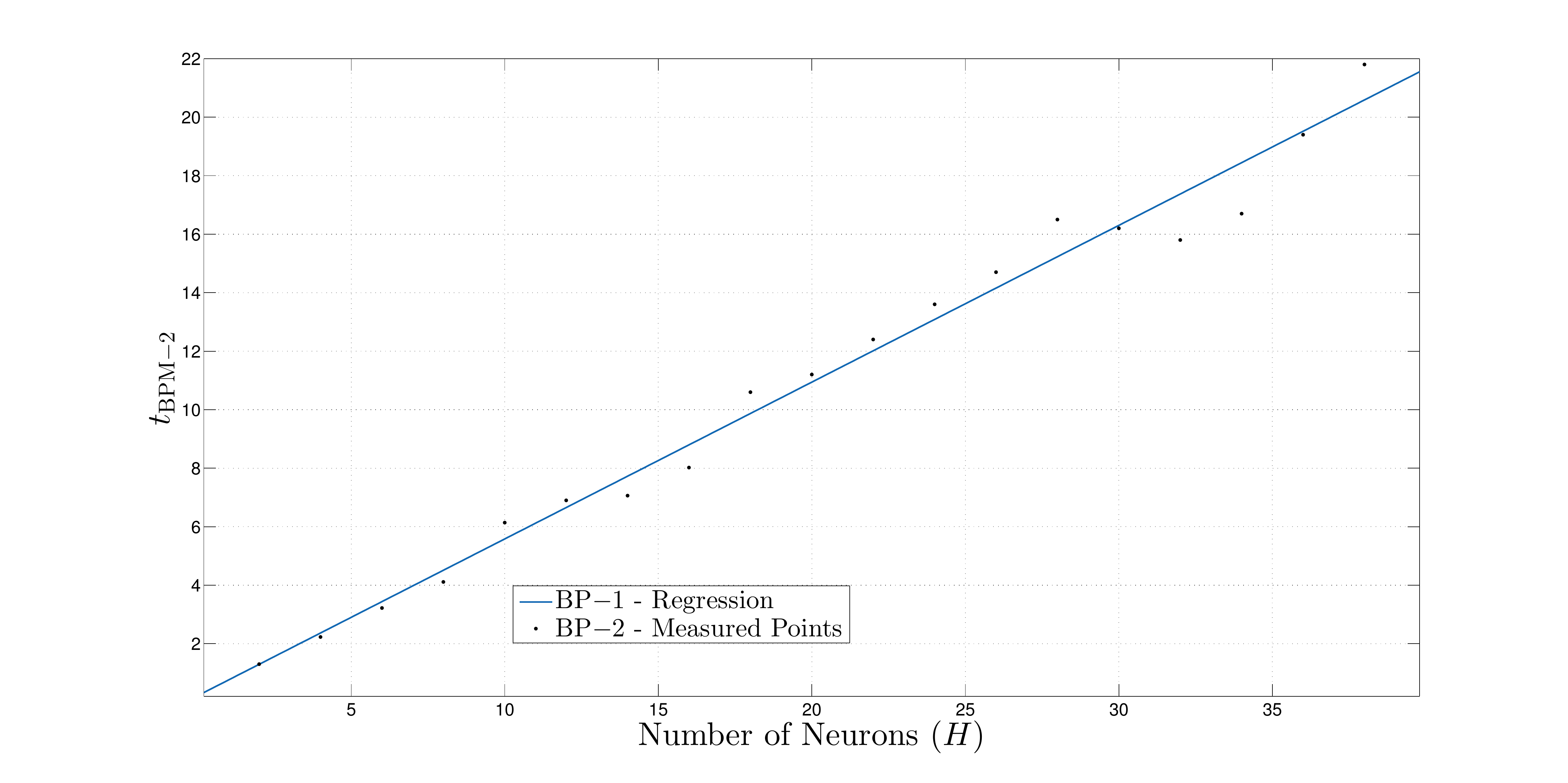}
				\label{FigResults4}
			}
			\caption{a) Execution time of $\text{BPM}-1$ by the number of neurons in the hidden layer, $H^1$. b) Execution time of $\text{BPM}-2$ by the number of neurons in the hidden layer, $H^1$.}
			\label{FigResults34}
		\end{center}
	\end{figure}

	An important result of these tests is the behavior of these fitted curves in Figures \ref{FigResults12} and \ref{FigResults34}. The processing time grows linearly with the number of neurons in the hidden layer ($H^1$). This result is quite significant since you can estimate if a certain $\mu$C can be used with this implementation and also if the network architecture that was chosen will fit or not in the $\mu$C. This result can be used as a reference by other groups that refute the usage of MLP-BP applications on $\mu$Cs.
	
	\begin{table*}[ht]
		\begin{center}
			\begin{tabular}{|c|c|c|c|c|c|}
				\hline
				$H^1$ & $t_{\text{FFM-}1} \, (\text{ms})$ & $t_{\text{FFM-}2}  \, (\text{ms}) $ & $t_{\text{EM}} \, (\text{ms})$ & $t_{\text{BPM-}1} \, (\text{ms})$ & $t_{\text{BPM-}2} \, (\text{ms})$ \\ \hline
				$2$  &           $0\text{.}59$           &            $0\text{.}31$            &         $0\text{.}06$          &           $0\text{.}46$           &           $1\text{.}30$           \\ \hline
				$4$  &           $4\text{.}28$           &            $0\text{.}61$            &         $0\text{.}06$          &           $0\text{.}66$           &           $2\text{.}23$           \\ \hline
				$6$  &           $6\text{.}21$           &            $0\text{.}77$            &         $0\text{.}06$          &           $0\text{.}88$           &           $3\text{.}22$           \\ \hline
				$8$  &           $8\text{.}11$           &            $1\text{.}00$            &         $0\text{.}06$          &           $1\text{.}11$           &           $4\text{.}11$           \\ \hline
				$10$  &          $10\text{.}50$           &            $1\text{.}82$            &         $0\text{.}06$          &           $1\text{.}70$           &           $6\text{.}14$           \\ \hline
				$12$  &          $12\text{.}30$           &            $2\text{.}02$            &         $0\text{.}06$          &           $1\text{.}91$           &           $6\text{.}90$           \\ \hline
				$14$  &          $14\text{.}50$           &            $1\text{.}49$            &         $0\text{.}06$          &           $1\text{.}80$           &           $7\text{.}06$           \\ \hline
				$16$  &          $16\text{.}30$           &            $1\text{.}72$            &         $0\text{.}06$          &           $2\text{.}03$           &           $8\text{.}02$           \\ \hline
				$18$  &          $18\text{.}60$           &            $2\text{.}57$            &         $0\text{.}06$          &           $2\text{.}65$           &          $10\text{.}60$           \\ \hline
				$20$  &          $20\text{.}70$           &            $2\text{.}71$            &         $0\text{.}06$          &           $2\text{.}47$           &          $11\text{.}20$           \\ \hline
				$22$  &          $22\text{.}60$           &            $2\text{.}95$            &         $0\text{.}06$          &           $3\text{.}07$           &          $12\text{.}40$           \\ \hline
				$24$  &          $25\text{.}20$           &            $3\text{.}15$            &         $0\text{.}06$          &           $3\text{.}36$           &          $13\text{.}60$           \\ \hline
				$26$  &          $27\text{.}30$           &            $3\text{.}31$            &         $0\text{.}06$          &           $3\text{.}71$           &          $14\text{.}70$           \\ \hline
				$28$  &          $29\text{.}30$           &            $3\text{.}51$            &         $0\text{.}06$          &           $3\text{.}95$           &          $16\text{.}50$           \\ \hline
				$30$  &          $30\text{.}90$           &            $3\text{.}67$            &         $0\text{.}06$          &           $4\text{.}03$           &          $16\text{.}20$           \\ \hline
				$32$  &          $32\text{.}90$           &            $3\text{.}76$            &         $0\text{.}06$          &           $3\text{.}83$           &          $15\text{.}80$           \\ \hline
				$34$  &          $34\text{.}60$           &            $3\text{.}36$            &         $0\text{.}06$          &           $4\text{.}05$           &          $16\text{.}70$           \\ \hline
				$36$  &          $36\text{.}50$           &            $4\text{.}22$            &         $0\text{.}06$          &           $4\text{.}82$           &          $19\text{.}40$           \\ \hline
				$38$  &          $38\text{.}50$           &            $4\text{.}41$            &         $0\text{.}06$          &           $5\text{.}17$           &          $21\text{.}80$           \\ \hline
			\end{tabular}
			\caption{Duration of execution for each module of the implementation of the MLP-BP with $P=H^0=2$, $M=H^2=1$ and sigmoid activation function.}
			\label{TabResults1}
		\end{center}
	\end{table*}
	
\begin{figure*} [ht]
		\begin{center}
			\includegraphics[width=.75\textwidth]{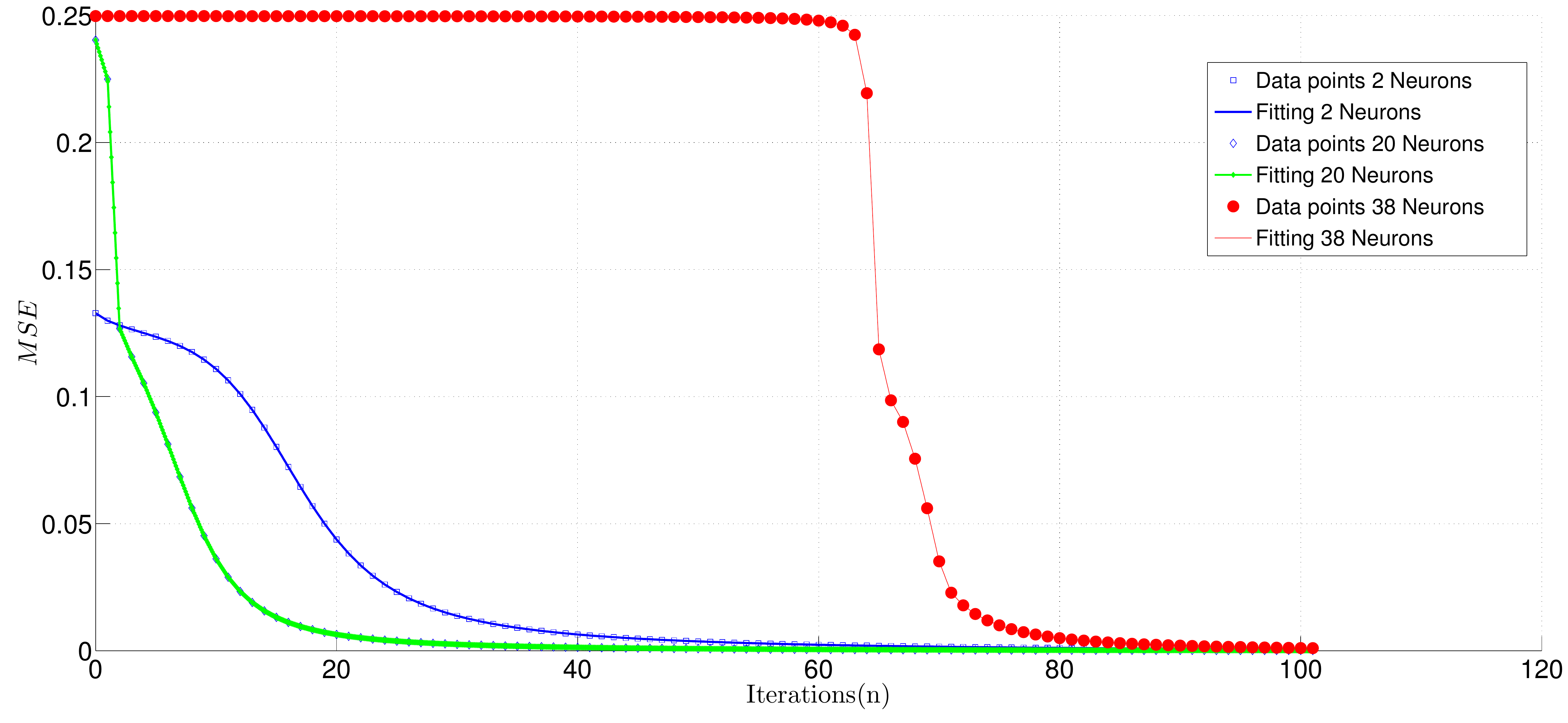}
			\caption{Mean Squared Error curve fitting comparison with $H^1=\{2, 20, 38\}$ neurons in the hidden layer.}
			\label{FigureMSE}
		\end{center}
	\end{figure*}

	Also we compared the MSE curve when varying the number of neurons in the hidden layer ($H^1$), as shown in Figure \ref{FigureMSE}. This shows that for 2 and 20 iterations the MSE was really close to zero, and that with 38 neurons, 80 iterations were needed. According to Table \ref{TabResults1} it took approximately 8.28 seconds to fully train the XOR case with 38 neurons.
	
	\subsection{Virtual Car-like Robot}
	
\begin{figure}[ht]
		\centering
		\includegraphics[width=0.75\linewidth]{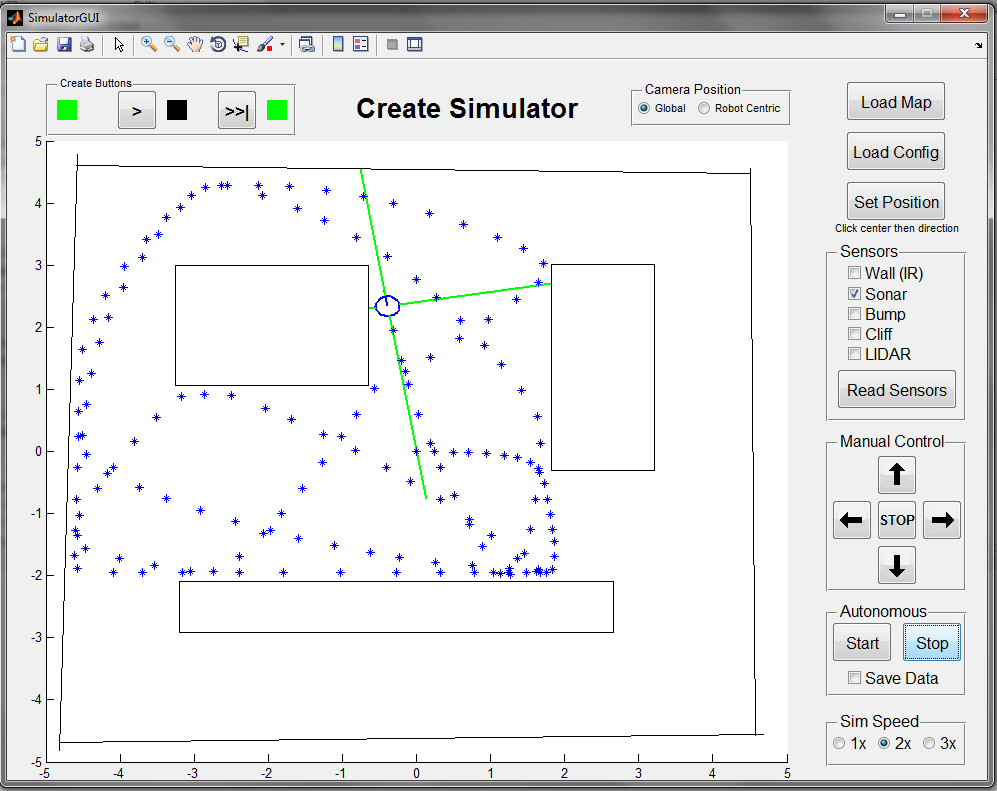}
		\caption{Virtual map with the resulting trajectory of the virtual car-like robot for the $1^{st}$ training dataset.}
		\label{fig:trajetoriacasosimples1}
	\end{figure}

	The first validation test-case showed small MSEs, close to $1\%$, with the architecture described in the methodology section, three inputs and two outputs with only five neurons in the hidden layer, ($H^1 = $). Since it was trained with a maximum distance of 1.0m, the robot shows a great behavior close to obstacles, but on big empty spaces with greater than 1.0m distances between obstacles the robot drifts and spins around the same spot until the test is restarted. This showed us that this training dataset was too simple and required more complex data. The test results for this dataset are shown in Figure \ref{fig:trajetoriacasosimples1}.
	
\begin{figure}[ht]
		\centering
		\includegraphics[width=0.75\linewidth]{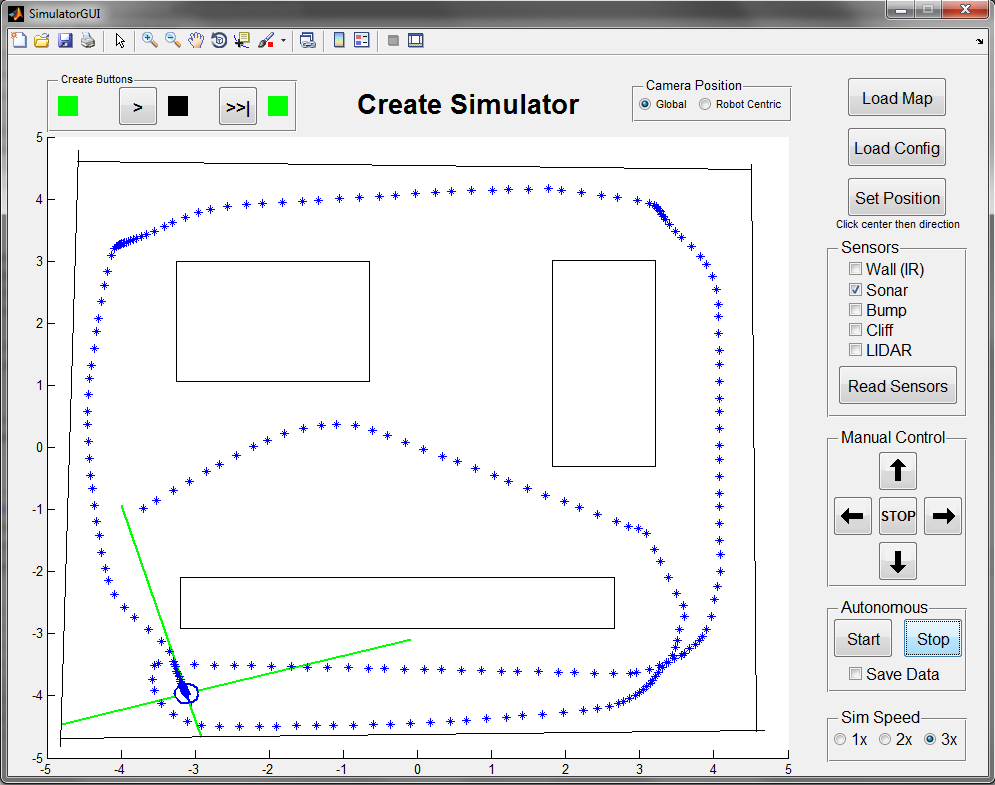}
		\caption{Virtual map with the resulting trajectory of the virtual car-like robot for the $2^{nd}$ training dataset.}
		\label{fig:trajetoriacasosimples2}
	\end{figure}

	The second dataset test showed much better results, with the virtual robot being able to react quickly to farther obstacles and also correcting its trajectory faster. The robot was able to run along the map borders smoothly and keep a safe distance between parallel obstacles on the center of the map. However, the virtual robot still collided with the obstacles and sometimes it would collide and drag it self along the borders of bigger obstacles, as shown in Figure \ref{fig:trajetoriacasosimples2}.
	
\begin{figure}[ht]
		\centering
		\includegraphics[width=0.75\linewidth]{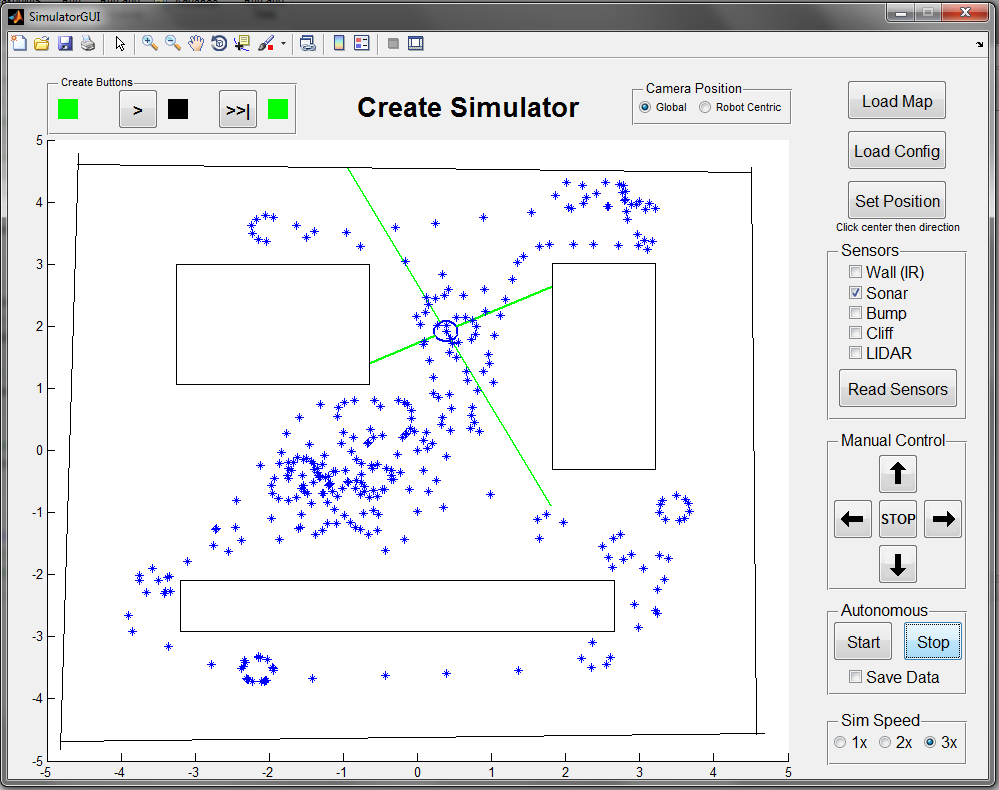}
		\caption{Virtual map with the resulting trajectory of the virtual car-like robot for the $3^{rd}$ training dataset.}
		\label{fig:trajetoriacasosimples3-2}
	\end{figure}

	The dataset for the third validation case resulted with a higher MSE of $2.4\%$. The car-like robot was faster than the previous two tests and was able to perform a fast reaction to abrupt changes in proximity to obstacles, as shown in Figure \ref{fig:trajetoriacasosimples3-2}. However, this dataset training made the robot show a behaviour that can be interpreted as overfitting, since sometimes it would react too quickly and start to spin around itself after detecting some obstacles ahead.
	
	The overall results shows that the datasets had not enough data and this required some more tests with even more virtual sensors. We included the angle between the Front-Sensor (FS) and the virtual horizontal axis of the test map. Also we performed training tests with two and three more hidden layers (L = {3,4}) and also hundreds of hidden neurons and thousands of iterations, none of which improved the results, actually preventing the Neural Network from reaching training error convergence lower than $20\%$.
	
	The generated binary code from the source in C for the best dataset ($3^{rd}$) had $3.07$KB of memory size, which amounts to $1.2\%$ of the $256$KB of program memory of the ATMega2560. This code size is quite compact compared to a similar model implementation seen in \cite{Ref4}. The same variables and testing criteria used for the XOR operation were used in this case, with 16Mhz of clock frequency.
	
	The timing results for the virtual car-like robot were expected, being smaller than the XOR operation that has a simpler architecture with less inputs and outputs ($H^0 = 1$ and $H^2 = 1$) and also a wide range for $H^1$, Table \ref{TabResults1}. Also, as this implementation does not perform online real-time embedded training, only the $\text{FFM-1}$, $\text{FFM-2}$ and $\text{FFM-3}$ modules were used, being $\text{FFM-2} and \text{FFM-3}$ for the cases with more than one hidden layer.
	
	The timing results for the first and second datasets were the same. This is expected, since the only factor to influentiate the execution time is size of the synaptic weight matrices $H^1$ and $H^2$ that depends only on the amount of inputs, neurons and the amount of samples. Since this was not a batch classification, but a single sample, online classification, the input size remained as $1$ and unchanged throughout all the three tests.
	
	The third case presented a fast timing result, but close to two times the timing results for the first and second cases. This is expected and simply reiterates what was shown in Figures \ref{FigResults12} and \ref{FigResults34}, that the execution time grows linearly with the hyperparameters or more specifically the amount of neurons for this implementation proposal.
	
	It is important to notice that the execution time results shown in Table \ref{tab:temposirobot} relate to what is found in the field's literature, that more hidden layers do not necessarily improves the training MSE and can actually make it worse, as seen in \cite{DBLP:journals/corr/abs-1709-02260}. In the cases were we trained the robot with more than a single hidden layer it's ability to avoid obstacles did not improve and the best MSE was slightly higher than $20\%$.
	
	\begin{table*}[ht]
		\begin{center}
			\begin{tabular}{|c|c|c|c|c|}
				\hline
				Datasets & $H^1$ & $t_{\text{FFM-}1} \, (\text{ms})$ & $t_{\text{FFM-}2}  \, (\text{ms}) $ & $t_{Total-\text{FFM-}k}$\\
				\hline			
				1 & 5 & $1.28$ & $0.52$ & $1.80$\\
				\hline
				2 & 5 & $1.28$ & $0.52$ & $1.80$\\
				\hline
				3 & 10 & $2.63$ & $0.88$ & $3.51$\\
				\hline
			\end{tabular}
		\end{center}
		\caption{Execution times measured with HIL for the implementation proposal in the virtual car-like robot validation.}
		\label{tab:temposirobot}
	\end{table*}
	
	\section{Comparison with the State-of-the-Art}
	
	The work McDanel et Al\cite{DBLP:journals/corr/abs-1709-02260} shows two implementations of an MLP-BP with single and dual hidden layers and their respective execution times for a classification with the results shown in Table \ref{tab:ebnn}. In \cite{gural19}, the authors were able to embed a full MNIST-10 classification model using CNNs under $2$KB of SRAM being used, also the inference times were in the order of 640 ms per input sample. 
	
	\begin{table*}[ht]
		\centering
		\begin{tabular}{|c|c|c|c|}
			\hline
			Reference & Model & Time (ms) & Memory (KB)\\ \hline
			McDanel et Al\cite{DBLP:journals/corr/abs-1709-02260} & MLP-1 &   17.35   &    14.73    \\ \hline
			McDanel et Al \cite{DBLP:journals/corr/abs-1709-02260} & MLP-2 &   9.17    &    13.53    \\ \hline
			Gural \& Murmann \cite{gural19} & AvgPool2x2 + CNN3x3 + MaxPool + MLP & 684 & 8.46 \\ \hline
			\textbf{Proposal} & MLP-2 & 69.88* & 6.67 \\ \hline
		\end{tabular}
		\caption{ANN results for the state-of-the-art works.*Time of inference, not taking the EM-k and BPM-k modules time into account.}
		\label{tab:ebnn}
	\end{table*}
	
	It's important to notice that not of the times presented in Table \ref{tab:ebnn} take only the inferencing time. The implementation for McDanel et Al \cite{DBLP:journals/corr/abs-1709-02260} trained the model offline using MatLab. For the Gural \& Murmann \cite{gural19} paper, the training was executed offline as well on a Jupyter Notebook running Tensorflow-Keras models and optimizations before porting the synaptic weights and kernels to the $\mu$C.
	
	It is possible to analyze this proposal, regarding how this system would behave if it had to support the same hyperparameters that some of these state-of-the-art applications have. The work presented by McDanel et Al\cite{DBLP:journals/corr/abs-1709-02260} uses an MLP ANN with a single layer with 100 artificial neurons. This same work also presents another implementation with a two-layer MLP with 200 artificial neurons. Analyzing the $t_{\text{FFM-}1}, t_{\text{FFM-}2}, t_{\text{BPM-}1}$ and $t_{\text{BPM-}2}$ fitted curves from Table \ref{TabResults1} we obtain four predictive equations that define how much time is required to process this MLP implementation.
	
	Equation \ref{eq:ffm1} shows us again that this works proposal has a linear relationship between inference time and hyperparameters, especifically artificial neurons. Evaluating this Equation \ref{eq:ffm1} with a number of 100 neurons it is observed $1.18 \text{s}$ of feedfowarding time for the first layer.
	
	\begin{equation}
		t_{\text{FFM-}1}\left(H^1\right) = 11.6 \times H^1 + 20.52
		\label{eq:ffm1}
    \end{equation}
    Taking in regard the same amount of neurons for the Equation \ref{eq:ffm2} we get $124.49 \text{ms}$ of feedforwarding time for the second layer. This amount to $1.30 \text{s}$ of total inferencing time with 100 neurons.
   	\begin{equation}
	    t_{\text{FFM-}2}\left(H^2\right) = 1.22 \times H^2 + 2.49
	    \label{eq:ffm2}
    \end{equation}
    If the embedded training is to be also considered with this amount of neurons, the time duration for this process increases to $1.44 \text{s}$, by evaluating the Equation \ref{eq:bpm1}.
   	\begin{equation}
	    t_{\text{BPM-}1}\left(H^2\right) = 1.41 \times H^2 + 2.72
	    \label{eq:bpm1}
    \end{equation}
    Also evaluating the Equation \ref{eq:bpm2} for the second layer, considering two hidden layers in training the time duration increases to $2.06 \text{s}$.
   	\begin{equation}
	    t_{\text{BPM-}2}\left(H^1\right) = 6.03 \times H^1 + 10.94
	    \label{eq:bpm2}
    \end{equation}
    This analysis shows that the modular implementation of a MLP here presented performs on compatible training and inferecing times.
	\section{Conclusions}
	
	This work presents an implementation proposal of an MLP artificial neural network with embedded BP training for an 8-bit $\mu$C. Results and implementation details were presented for this proposal for an ATMega-$2560$ $\mu$Cs. Also, the validation results of the embedding of this proposal were presented using a HIL simulation strategy. Finally, the results show that the execution times and memory occupation of the implementation were compatible with application requirements seen in the industry, seen that these requirements fall under hundreds of milliseconds.

\bibliographystyle{spmpsci}  
\bibliography{PaperMain}

\end{document}